\documentclass[sigconf]{acmart} 

\usepackage{amsmath, amsfonts, array, arydshln}
\usepackage{algorithmic}
\usepackage{enumitem}
\usepackage{graphicx}
\usepackage{textcomp}
\usepackage{array}
\usepackage{verbatim}
\usepackage{xcolor}
\usepackage{caption}
\usepackage{enumitem}
\usepackage{float}
\usepackage{tabularx}
\usepackage{longtable}

\newcolumntype{Y}{>{\raggedright\arraybackslash}X}

\copyrightyear{2024}
\acmYear{2024}
\setcopyright{rightsretained}
\acmConference[UIST '24]{The 37th Annual ACM Symposium on User Interface Software and Technology}{October 13--16, 2024}{Pittsburgh, PA, USA}
\acmBooktitle{The 37th Annual ACM Symposium on User Interface Software and Technology (UIST '24), October 13--16, 2024, Pittsburgh, PA, USA}
\acmDOI{10.1145/3654777.3676462}
\acmISBN{979-8-4007-0628-8/24/10}

\begin{document}

\title{FathomGPT: A Natural Language Interface for Interactively Exploring Ocean Science Data}

\author{Nabin Khanal$^*$}
\thanks{$^*$Both authors contributed equally to this research.}
\email{khanaln@purdue.edu}
\orcid{0009-0007-1972-7022}
\affiliation{%
  \institution{Purdue University}
  \city{West Lafayette}
  \state{Indiana}
  \country{USA}
}

\author{Chun Meng Yu$^*$}
\email{yu1327@purdue.edu}
\orcid{0009-0000-1324-044X}
\affiliation{%
  \institution{Purdue University}
  \city{West Lafayette}
  \state{Indiana}
  \country{USA}
}

\author{Jui-Cheng Chiu}
\email{chiu119@purdue.edu}
\orcid{0009-0000-4180-3852}
\affiliation{%
  \institution{Purdue University}
  \city{West Lafayette}
  \state{Indiana}
  \country{USA}
}

\author{Anav Chaudhary}
\email{chaudh75@purdue.edu}
\orcid{0000-0003-4917-4848}
\affiliation{%
  \institution{Purdue University}
  \city{West Lafayette}
  \state{Indiana}
  \country{USA}
}

\author{Ziyue Zhang}
\email{zhan5079@purdue.edu}
\orcid{0009-0000-7379-9317}
\affiliation{%
  \institution{Purdue University}
  \city{West Lafayette}
  \state{Indiana}
  \country{USA}
}

\author{Kakani Katija}
\email{kakani@mbari.org}
\orcid{0000-0002-7249-0147}
\affiliation{%
  \institution{MBARI}
  \city{Moss Landing}
  \state{California}
  \country{USA}
}

\author{Angus G. Forbes}
\email{agforbes@purdue.edu}
\orcid{0000-0002-8700-7795}
\affiliation{%
  \institution{Purdue University}
  \city{West Lafayette}
  \state{Indiana}
  \country{USA}
}

\begin{abstract}
  We introduce FathomGPT, an open source system for the interactive investigation of ocean science data via a natural language interface. FathomGPT was developed in close collaboration with marine scientists to enable researchers to explore and analyze the FathomNet image database. FathomGPT provides a custom information retrieval pipeline that leverages OpenAI’s large language models to enable: the creation of complex queries to retrieve images, taxonomic information, and scientific measurements; mapping common names and morphological features to scientific names; generating interactive charts on demand; and searching by image or specified patterns within an image. In designing FathomGPT, particular emphasis was placed on enhancing the user's experience by facilitating free-form exploration and optimizing response times. We present an architectural overview and implementation details of  FathomGPT, along with a series of ablation studies that demonstrate the effectiveness of our approach to name resolution, fine tuning, and prompt modification. We also present usage scenarios of interactive data exploration sessions and document feedback from ocean scientists and machine learning experts.
\end{abstract}


\maketitle

\captionsetup[table]{skip=0pt} 

\begin{figure*}
    \centering
 \includegraphics[width=\textwidth]{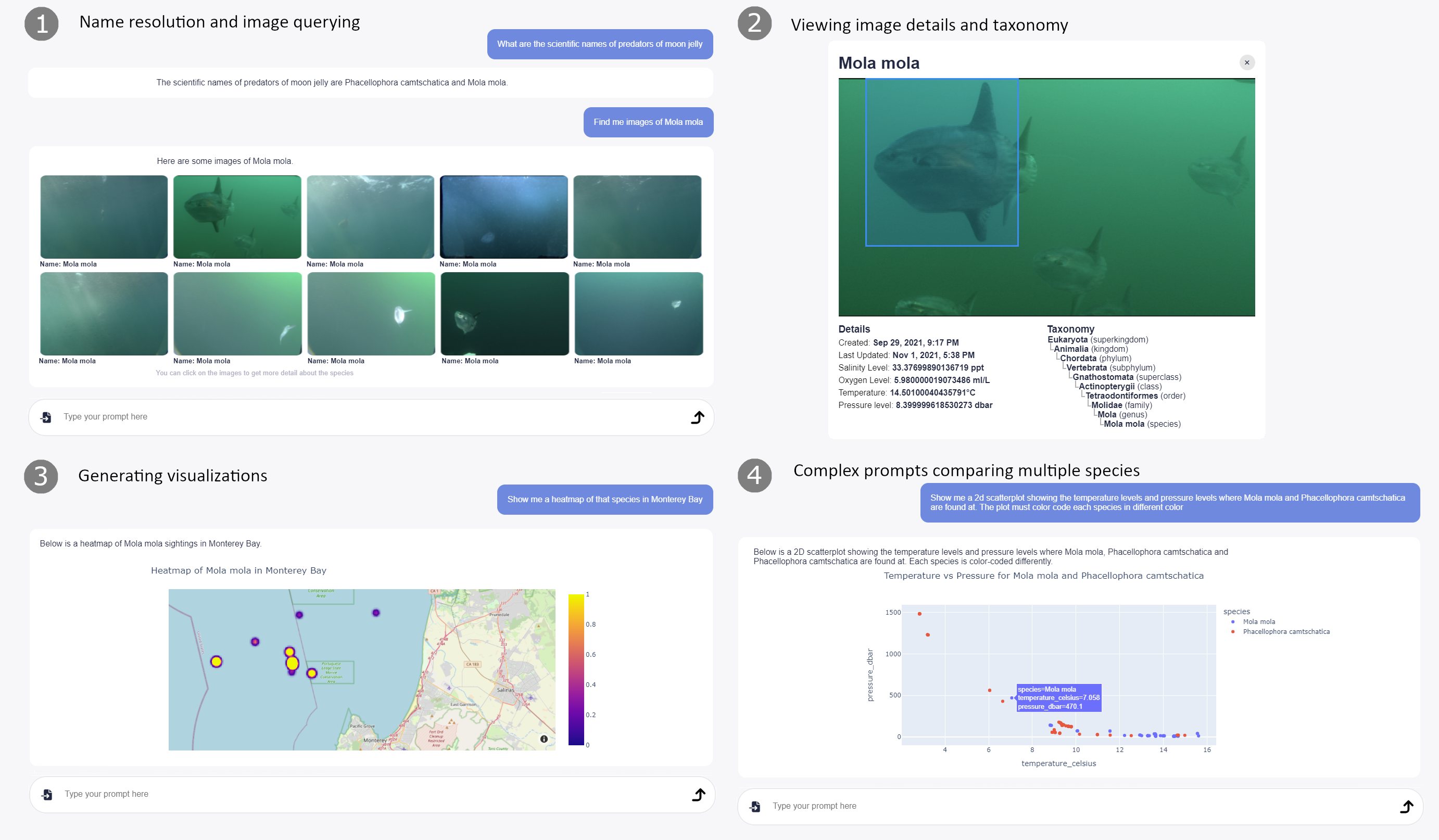}
  \caption{This figure shows four successive screenshots from a single session using the FathomGPT natural language interface for exploring ocean science data. (1) The user asks FathomGPT to return the scientific names for predators of the species commonly known as ``moon jelly,'' and then asks to see images of Mola mola, one of the predators. (2) The user then clicks on an image to see more details about the taxonomy of the species, as well as to learn where and when that particular image was taken and what scientific measurements were gathered at that time. (3) The user then asks FathomGPT to generate an interactive heatmap showing other observations of Mola mola in the Monterey Bay region. (4) Finally, the user asks FathomGPT to generate a detailed scatterplot comparing the temperature and pressure levels at the locations where Mola mola (``ocean sunfish'') and another predator Phacellophara camtschatica (``egg-yolk jellyfish'') have been observed. FathomGPT facilitates free-form exploration of ocean science data from multiple starting points to support a wide range of research inquiries.}
  \label{fig:teaser}
\end{figure*}

\section{Introduction}

The ocean is experiencing unprecedented rapid change, and visually monitoring marine biota at the spatiotemporal scales needed for responsible stewardship is a formidable task. The volume and rate of this required data collection rapidly outpaces our abilities to process and analyze them. However, recent advances in machine learning have enabled fast, sophisticated analysis of visual data. FathomNet~\cite{katija2022fathomnet} is an open-source image database that standardizes and aggregates expertly curated labeled data, including imagery of marine animals, underwater equipment, debris, among other concepts, and encourages contributions from distributed data sources. Additionally, FathomNet data can be used to train and deploy models on other institutional videos to reduce annotation effort and enable automated tracking of underwater concepts when integrated with robotic vehicles, accelerating the processing of visual data to achieve a healthy and sustainable global ocean. 

FathomNet currently hosts 109,871 images containing 296,795 individual species, and each month over 2,000 additional images are ingested. FathomNet is publicly available via both an interactive website and a REST API, and more advanced users can interact with the FathomNet database directly using SQL queries. However, despite the increasing richness of this dataset and the increasing need for researchers to make use of machine learning tools to annotate and analyze images, it remains difficult to support a range of tasks useful for the intended audience, which includes ocean scientists, policy makers, and ocean enthusiasts. In a 2023 survey, Crosby et al.~\cite{crosby2023designing} delineate core issues that prevent researchers and other users from incorporating scientific databases into their scientific workflows, which include: challenges with data sharing, a lack of expertise in machine learning technologies, and a lack of accessible tools and interfaces. 

We introduce FathomGPT to address these issues, to promote database accessibility, and to support a wide range of data explorations involving retrieving, visualizing, and analyzing ocean images, species annotations, taxonomic information, and scientific measurements. FathomGPT introduces a natural language interface that leverages large language models (LLMs) to enable different exploration pathways and to facilitate a range of scientific analysis tasks. We worked closely with our marine science collaborators to identify the core tasks that FathomNet is intended to support, and we participated in a series of workshops to elicit additional usage scenarios for interacting with the FathomNet database (Section~\ref{Sec:UsageScenarios}). Based on this feedback, we designed FathomGPT to enable users to freely write natural language prompts in order to make it easy to: ask general knowledge questions about marine life and ocean ecosystems; look up the scientific name of species via their colloquial name or by providing descriptions of their morphology, their color, their predator/prey relationships, or their habitat; craft complex database queries that enable users to make comparisons between species, to understand changes over time, and to reason about relationships between scientific measurements and species observations; review ``data cards'' consisting of annotated images, taxonomic information, and scientific measurements; search for species observations by uploading an image or by interactively selecting/segmenting a specific pattern within the uploaded image; and generate custom interactive charts.

During each FathomGPT session, our system retains a memory of previous user prompts, so that a user can, for example, refine or broaden a query, modify a visualization, or make additional inquiries regarding particular concepts. As with other LLM interfaces, the interaction style sometimes resembles a conversation with an intelligent agent. Our contributions include various technical innovations (described in Section~\ref{Sec:FathomGPT}) that facilitate robust interactions:

\begin{itemize}[leftmargin=1.2em,parsep=0em,partopsep=0em,topsep=0.3em,itemsep=0.3em]

\item 
 We present an efficient architecture (see \autoref{fig:Pipeline}) optimized to reduce both overall latency and minimize the tokens used so that our system can retrieve information from a relational database within a few seconds in most cases (see \autoref{table:query_performance}).

\item We introduce a prompt modification system that integrates relevant context from the conversation to generate effective SQL queries or python code while reducing the token count (see \autoref{table:prompt_modification}).

\item We show that the use of multiple specialized models to perform text-to-SQL generation tasks supporting different output types improves accuracy compared to using a single ``monolithic'' text-to-SQL generation model (see Section~\ref{SubSec:Text2SQL}).

\item
We present a method for identifying species based on various features and for resolving common names by leveraging GPT to generate structured knowledge graphs from Wikipedia text and taxonomic databases, implementing a graph alignment technique to match subject-relation-object triples between the knowledge graph and user prompts (see Section~\ref{SubSec:NameResolution}).

\item 
We present an interactive tool for highlighting patterns on an image and using those patterns to retrieve images with similar patterns from the database (see Section~\ref{SubSec:PatternAnalysis}).

\end{itemize}

\noindent In addition to enabling an effective natural language interface to support the exploration and analysis of data housed in the FathomNet database, we believe our approach could be applied to other scientific databases, accelerating research tasks and promoting accessibility~\cite{bell2022low}.


\section{Related Work}

\subsection{LLM-powered user interactions}

Many recent projects explore the use of large language models (LLMs) to support innovative interfaces. For example, Promptify~\cite{brade2023promptify} utilizes a suggestion engine powered by LLMs to help users quickly explore and craft diverse prompts for text-to-image generative models, allowing users to create appealing images on their first attempt while requiring significantly less cognitive load. Graphologue~\cite{Jiang2023Graphologue} provides a means to identify ``long-winded responses,'' which can be one of the issues faced by users working with LLMs, and proposes a method of crafting interactive graphs that break responses into smaller node-link graphical chunks that are much easier to digest. Chen et al.~\cite{Chen2023Gap} present an interaction method between a user and an LLM in which the model crafts and retains a personalized memory of the user, enabling better context awareness and producing more meaningful responses. GenAssist~\cite{huh2023genassist} is a system that makes text-to-image generation accessible for blind and low-vision creators. The system uses LLMs to generate visual questions and uses specialized vision-language models to extract answers and summarize results, assisting creators in verifying whether or not generated image candidates accurately followed their initial prompt. 

The FathomGPT system provides a natural language interface that enables users to conduct free-form explorations of ocean science data and also makes use of vision-language models to search for specified patterns and images within the database.

\subsection{Scientific information retrieval using LLMs}
The rapid growth in large language models (LLMs) featuring task-agnostic architectures and pre-training has led to their use in a wide range of applications~\cite{vaswani2017attention, brown2020language, bommasani2021opportunities}. LLMs such as OpenAI's GPT models have shown remarkable performance across different domains of natural language processing tasks including translation, summarization, question answering, and generation~\cite{devlin2018bert, brown2020language, bommasani2021opportunities, achiam2023gpt}. Our system leverages OpenAI's GPT models to connect to external tools, to generate complex SQL queries, and to generate custom code that retrieves and visualizes information from the open-source FathomNet ocean image database~\cite{katija2022fathomnet}.


Recent work in geospatial data analysis demonstrates the effectiveness of using LLMs to translate natural language prompts into SQL queries, facilitating access to spatial databases~\cite{jiang2024chatgpt}. Another recent project uses an LLM model to target medical decision-making, showing increases in speed and accuracy on a range of retrieval tasks~\cite{ke2024development}. These applications underscore the transformative impact of LLMs in information retrieval.

One popular approach to generating and presenting data using LLMs is retrieval-augmented generation, or RAG~\cite{gao2023retrieval, lewis2020retrieval}. RAGs incorporate data from external sources--- such as structured databases, unstructured text from the web, scientific articles, PDF documents, and other domain-specific data--- in order to generate information that enhances the accuracy and credibility of the generated responses~\cite{lewis2020retrieval}. The FathomGPT pipeline incorporates prompt-to-SQL generation to quickly retrieve results from the FathomNet database.



\subsection{Generating SQL queries} 
Generating SQL queries from natural language prompts is often framed as a sequence-to-sequence problem, leveraging LSTMs or transform\-er-based architectures for supervised learning~\cite{sutskever2014sequence, zhong2017seq2SQL, vaswani2017attention, hwang2019comprehensive}. More recently, LLMs have shown that they are effective in converting text to SQL~\cite{raffel2020exploring}. The use of LLMs in combination with particular prompting strategies currently lead text-to-SQL benchmarks~\cite{li2024codes, yu2018spider, li2024can}. Recent research investigates how best to design prompt engineering strategies and how best to represent questions and database schemas in order to produce good results when interacting with LLMs~\cite{gao2023texttoSQL}. 

Effective prompting can lead to drastic performance improvements~\cite{reynolds2021prompt}. Providing the model with demonstrations is known as in-context learning~\cite{brown2020language}. In this approach, exemplars are provided so that an input prompt returns an ideal response from the model. Based on the number of examples provided, the prompting strategy can be classified as zero-shot, one-shot, or few-shot. Another approach to prompting involves providing Chain-of-Thought demonstrations~\cite{wei2022chain} that ask the model to generate a series of intermediate reasoning steps, which can improve performance on a range of reasoning tasks. Yet another approach called ReAct interweaves a series of verbal reasoning and task-related actions, enhancing the LLM's ability to develop and adjust reasoning strategies by incorporating new information~\cite{yao2022react}. 

FathomNet uses a few-shot in-context learning approach to generate accurate results and reduce the overall response time. This provides a balanced approach that emphasizes engaging user interactions and optimizes the way in which prompts are modified to require fewer tokens while still producing meaningful results.




\subsection{Generating visualizations}
A number of recent projects have shown promising results in using LLMs to generate data visualizations~\cite{vazquez2024llms}. For example, LIDA generates grammar-agnostic visualizations using 4 modules consisting of a summarizer, a goal explorer, a visgenerator, and an infographer~\cite{dibia2023lida}. Similarly, ChartGPT decomposes the visualization generation process into a step-by-step reasoning pipeline~\cite{chartGPT}. FathomGPT generates custom code using the Plotly Open Source Graphing Library~\cite{plotly} to quickly create custom interactive visualizations requested by the user.

\subsection{Name resolution with knowledge graphs}
A structured knowledge base, or knowledge graph (KG), contains semantic triples consisting of a subject, relation, and object that map an entity to its characteristics or to other entities. There are many knowledge bases available--- including Wikidata and Freebase~\cite{yao2014information}--- that can be used for question-answering tasks, such as retrieving data using natural language prompts from a wide variety of databases~\cite{park2021knowledge, sun2021covid}.

Name resolution and concept resolution map common names and descriptions to a list of scientific names. For ocean science research, name resolution is an important task since species are known by many different names across the world, yet in scientific databases and in the research literature references to species are standardized using scientific names. Additionally, it can be useful to resolve species \textit{characteristics} to scientific names, even when a common name of a creature is not known. 

The KG required for resolution needs to contain mappings between the scientific name and the characteristics of a species, such as their coloring, their habitat, or list of their predators. The available structured knowledge bases containing scientific data, such as Wikidata and WoRMS~\cite{costello2013global}, have a limited amount of data compared to the plaintext data available on Wikipedia. For example, they do not provide the color of the creature as a separate field. Our work focuses on leveraging LLMs to generate KGs from Wikipedia pages using prompt engineering to extract species characteristics from the plain-text data.


The user prompt also needs to be structured, and we use LLMs to convert the prompt into a semantic triple, similar to prior work that explores aligning natural language prompts with graph databases \cite{liang2024aligning}. Unlike previous work that trained Memory Networks for question answering tasks~\cite{bordes2015large}, FathomGPT does not require any additional training. We instead use a graph alignment method similar to the entity linking solution used for the Relation Extraction model by the LAMA probe~\cite{petroni2019language}, which performs string matching between subject-relation-object triples.

\subsection{Pattern analysis}
Marine biodiversity often presents distinctive patterns and morphological features~\cite{Baldwin2013Phylogenetic}, which can be used to identify and classify different species. Analyses of these patterns can shed light on genetic relationships within and evolutionary relationships between species~\cite{Varas2019Genetic}. Differences in patterns may indicate ecological preferences and adaptive capabilities, and species within the same habitat may have unique visual characteristics that are affected by the environment~\cite{Verberk2012General}. Additionally, exterior features within species can vary significantly when living in different habitats~\cite{Roches2017Ecological}. By examining patterns within different populations living in different places, we can elucidate the adaptive variation of these changes and their implications for species resilience and conservation. 


Pattern analysis methods support species and specimen identification, particularly for species with distinctive patterns. Stewart et al.~\cite{stewart2021animal} highlight the advantages of AI-based approaches in identifying individual animals as opposed to relying on human annotation. Furthermore, multiple studies have been conducted for land-based species with distinctive patterns. Parham et al.~\cite{parham2018animal} utilize AI-based approaches to annotating species in an image, targeting zebras, giraffes, and sea turtles, and Crall et al.~\cite{Crall2013Hotspotter} present a method of identifying individual animals of a species through both sequential and nearest-neighbor approaches using patterns as the discriminating factor. However, natural language representations of these patterns are relatively unexplored. Pattern representations exist in high dimensional feature spaces that discourage direct interaction with the pattern itself. Currently, to the best of our knowledge, LLMs are not able to effectively describe the patterns found in marine life through prompt conditioning and fine-tuning. 

FathomGPT introduces an interface for interactively guiding the user to highlight patterns on an image, and then to use those patterns to retrieve images of species with similar markings. Since human-annotator pipelines for marine life often depend on features such as shape, color, pose, and patterns, a natural language-based tool for interacting with pattern representations can further improve both database lookup and image annotation tasks, as well as improve current image captioning pipelines.

\section{The FathomGPT Platform} 
\label{Sec:FathomGPT}

FathomGPT was designed in close collaboration with our ocean science partners, and during the development process we also solicited feedback from museum installation specialists, machine learning experts, and user experience designers. While our primary audience is ocean scientists who are familiar with the FathomNet database, we focused extensively on the design of the user experience to make sure that interactions with FathomGPT were engaging and useful to a wide range of users. In particular, we were highly cognizant of the importance of optimizing response times. FathomGPT features a chatbot-like experience that can be used by users to retrieve and visualize information from the database, and a core design goal was to make sure users could interact with the database in a fast, reliable, and seamless manner. The system is designed to generate a response to a user’s query quickly, achieving a maximum response time of 30 seconds even for very complex requests (and most often under 5 seconds). Table ~\ref{table:query_performance} shows the response time the system takes for prompts of increasing complexity. After every prompt and between every intermediate processing step, Our interface also provides the user with information about what stage of the process the system is at currently.

\begin{figure}
    \centering
    \includegraphics[scale=0.345]{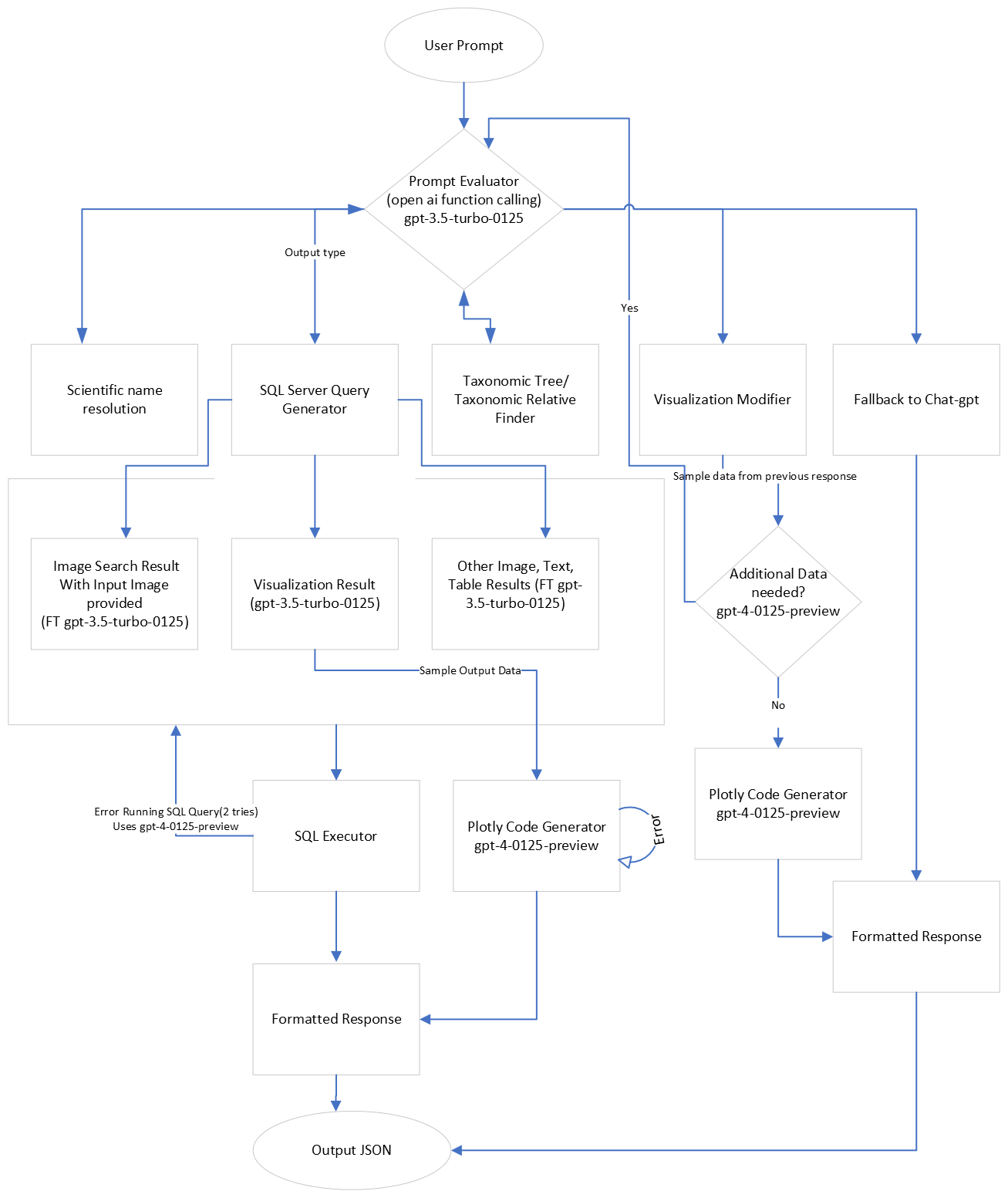} \\
    \caption{The FathomGPT pipeline. Here we show a high-level schematic of how an input prompt is processed by the system to produce a JSON output response, which is sent to the frontend webpage to be rendered.}
    \label{fig:Pipeline}
\end{figure}



\subsection{The prompt evaluator}
\label{subsec:promptevaluator}
At the core of the FathomGPT pipeline (Figure~\ref{fig:Pipeline}), we use an LLM that is able to determine how best to create meaningful responses to different types of user prompts, making use of the function calling features available in gpt-3.5-turbo-0125~\cite{openai2024api}. We provide the LLM with five FathomGPT prompt processing functions--- related to name resolution, query generation, taxonomic data, visualization code, and general information--- along with their description, the required input, and expected output type. 

During conversations involving successive prompts, this ``prompt evaluator'' model is the only part of the system that has access to the entire context from the user’s previous conversation. The model is instructed to reformat the user prompt, modifying it to include relevant aspects of the previous context, and only then pass this modified prompt to the appropriate functions. For example:
\begin{description}
    \item[Prompt 1:] ``Show me images of \textit{Merluccius productus}.''
    \item[Prompt 2:] ``What is the average temperature where that species is found according to the database?''
\end{description}
Here, despite the user's referential use of ``that'' in the second prompt, the prompt evaluator intelligently transforms it before passing it to other functions:
\begin{description}
    \item[Modified Prompt:] ``What is the average temperature where \textit{Merluccius productus} is found according to the database?''
\end{description}
(See Table~\ref{table:prompt_modification} for additional prompt modification examples.) 

The LLM’s remarkable ability to process text effectively will simplify all other processing steps following this step, as we now no longer need to pass the previous conversation to the LLM in downstream function calls. We use gpt-3.5-turbo-0125 instead of gpt-4-turbo to reduce the time it takes to execute an OpenAI API call~\cite{OpenAI2024GPT35Turbo, OpenAI2024GPT45Turbo}. However, the gpt-3.5-turbo-0125 model has a context length of 16k, which limits the length of the conversation that FathomGPT can have. Therefore, to preserve the tokens that can be passed for a conversation, this model is inputted with only the elements from the previous messages that might be needed to understand the previous context (see Section~\ref{SubSec:Ablation2}).


Once the prompt evaluator determines which next function is appropriate for the prompt, that function is run. If another step is needed for processing the input prompt, control is sent back to the prompt evaluator, along with the output from the most recent function. Otherwise, it is formatted and sent to the frontend. 

To prevent the prompt evaluator from falling into a loop when a function does not return the expected output, there is a limit on how many functions can be called by the evaluator for any single request. Within the function, we include error feedback functionality that pass any errors to the GPT-4 model to attempt to resolve them (see Section~\ref{subSubSec:Robustness}).

\begin{table}[t]
\centering
\caption{Query response times}
\begin{tabular}{p{6.75cm}|p{1cm}}
\hline
\textbf{Prompt} & \textbf{Time} \\ \hline
What color are aurelia aurita? & 2.53s \\ \hline
What is the scientific name of moon jellyfish? & 2.69s \\ \hline
Generate me a list of top 20 species from the database with their count. & 3.04s \\ \hline
Find me images of Aurelia aurita. & 4.62s \\ \hline
Find me images of moon jelly in Monterey bay and depth less than 5k meters. & 5.19s \\ \hline
What are the scientific names of orange creatures? & 7.33s \\ \hline
Display a bar chart showing the distribution of all species in Monterey Bay, categorized by the salinity level they are found in & 13.24s \\ \hline
Plot all the places where rockfish and Pacific hake are found in a map. The points that you show in the map should be color coded differently for each species. Also, make the size of the circle of each data point depend on the depth the species are found in. & 13.3s \\ \hline
Find me similar looking images from the database. & 15.59s \\ \hline
Display a bar chart showing the distribution of all orange creatures in Monterey Bay, categorized by the salinity level they are found in. & 23.85s \\ \hline

\end{tabular}
\label{table:query_performance}
\end{table}

\begin{table*}[ht]
\centering
\caption{Examples of how our prompt modification approach augments a user's prompt to add previous context prior to it being passed into any of the various functions that are invoked by the prompt evaluator.}
\begin{tabular}{p{5.6cm}|p{5.6cm}|p{5.5cm}}
\hline
\textbf{Prompt 1} & \textbf{Prompt 2} & \textbf{Modified prompt} \\ \hline
Display a heatmap of all species in Monterey Bay & Add the depth data so that when I hover over the data point, I can view it & Display a heatmap of all species in Monterey Bay with depth data included for viewing upon hovering over the data point \\ \hline
Find me images of Aurelia aurita & Show me a box plot of it with the data about the salinity level it is found in & Show me a box plot of Aurelia aurita with the data about the salinity level it is found in \\ \hline
Show a bar chart showing the depth level at which Aurelia aurita were found & Generate another for Bathochordaeus stygius & Show a bar chart showing the depth level at which Bathochordaeus stygius were found \\ \hline
\end{tabular}
\label{table:prompt_modification}
\end{table*}

\subsection{Name resolution}
\label{SubSec:NameResolution}

Compared to other methods for name resolution, such as querying the World Register of Marine Species (WoRMS) or using GPT-4 directly, our method expands the coverage of the common name to scientific name mapping from WoRMS by including names from Wikipedia, and at the same time focuses the results to only include concepts that are currently available in the ever-growing FathomNet database. Our method also supports name resolution using descriptions (e.g., ``tentacles'', ``orange creatures''), a feature that is not available in WoRMS and which will cause GPT-4 to return results that are not available as FathomNet concepts. Even when using the recently released GPT-4o and explicitly limiting it to only output FathomNet concepts, we found that this newer model would frequently generate hallucinations (see Section~\ref{SubSec:EvalNameResolution}).

Our name resolution process first attempts to map the common name to the scientific name using a pre-processed JSON dictionary. The data is pulled from Wikipedia and the WoRMS database. If the prompt contains a common name, we normalize the string and return the corresponding scientific names. 

\begin{figure}[t]
    \centering
    \includegraphics[scale=0.42]{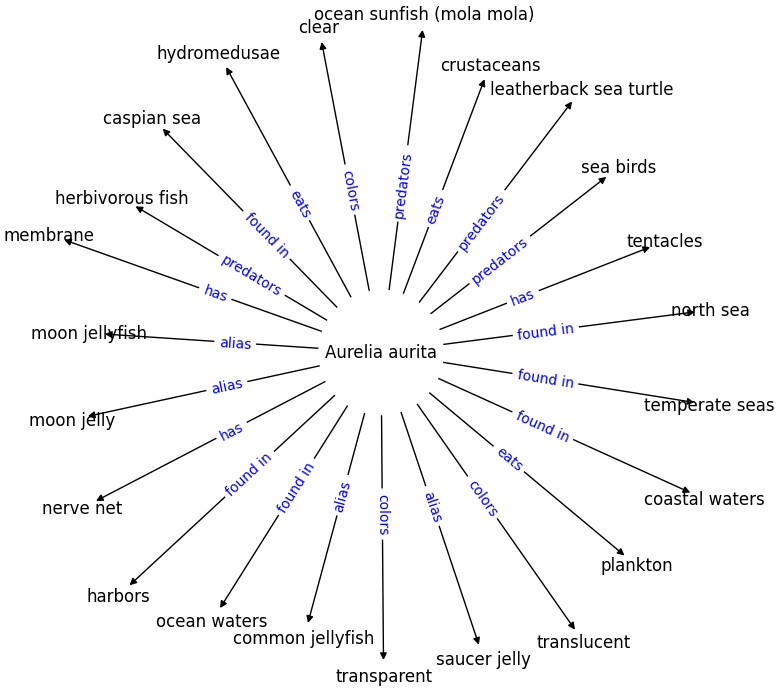} \\
    \caption{Example of a Species KG. Here we show the KG containing various characteristics for the species Aurelia aurita, including its common names, morphological features, colors, habitats, diet, and predators.}
    \label{fig:KG-example}
\end{figure}

If the prompt is more complex and requires processing features (rather than common names), then we utilize knowledge graph name resolution. This is our main method for name resolution involving creature descriptions. We create knowledge graphs from Wikipedia for each available creature, which we refer to as the ``Species KGs'' (see Figure~\ref{fig:KG-example}). 
We extract a subject-relation-object triple from the description of the species in the user prompt, which we refer to as the ``Prompt KG'', and we perform a graph alignment of the Prompt KG and the Species KGs to obtain the list of scientific names for creatures that match the prompt.

Each Species KG maps a scientific name to a list of its characteristics (morphology, color, predator/prey relations, environment). We generate the Species KG by providing GPT-3.5 with the text from Wikipedia's entry for the creature and specific instructions to extract various characteristics for each concept, including ``aliases'', ``body parts'', ``colors'', ``predators'', ``diet'', ``environment'', and ``descriptors''. This returns the corresponding characteristics extracted from the Wikipedia text as a JSON object (e.g. \textit{\{“colors”: [“orange”, “red”]\}}). We create a pre-processed list of Species KGs for each type of characteristic for every species and store it as a JSON file.

\begin{figure}[t!]
    \centering
    \includegraphics[scale=0.37]{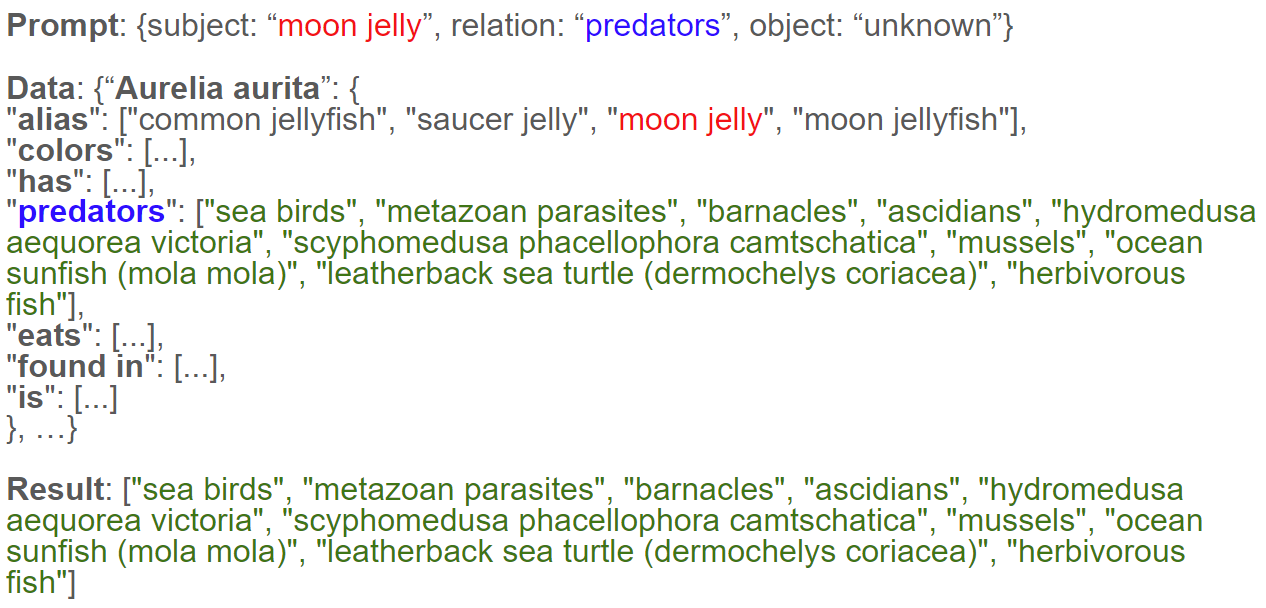} \\
    \caption{An example of KG subject matching. In the input prompt KG, the subject “moon jelly” matches the alias of Aurelia aurita in the species KG. The relation “predators” matches the same relation in the species KG. The result is the list of terms indexed by \textit{Data: [“Aurelia aurita”][“predators”]}.}
    \label{fig:KG-alignment-subject}
\end{figure}

\begin{figure}[h!]
    \centering
    \includegraphics[scale=0.31]{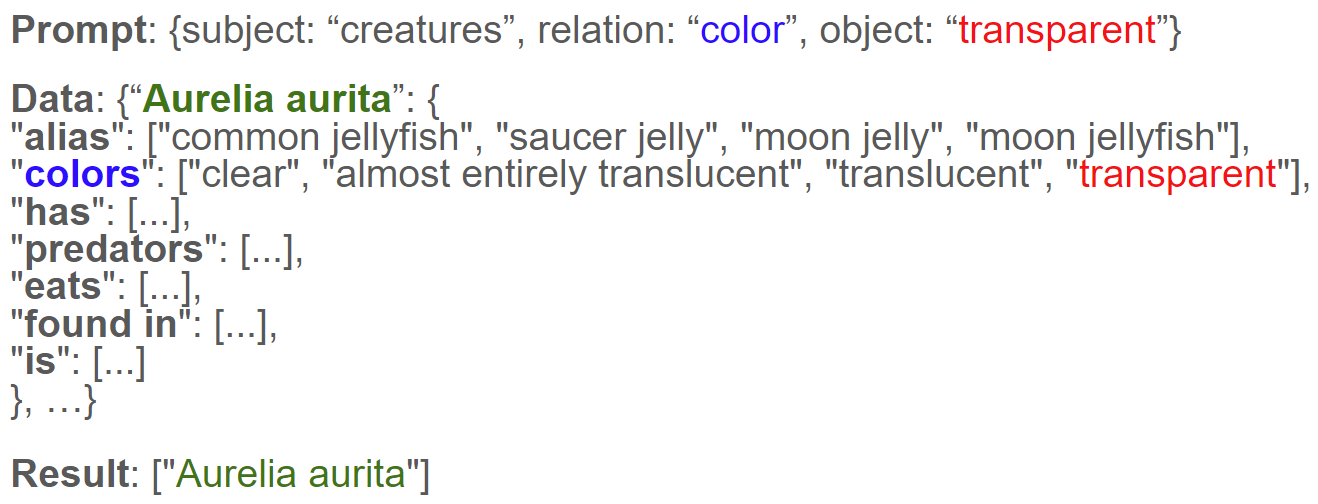} \\
    \caption{An example of KG object matching. In the input prompt KG, the relation “color” matches the same relation in the species KG and the object “transparent” matches one of the colors of Aurelia aurita in the species KG. The result would be the parent concept “Aurelia aurita”.}
    \label{fig:KG-alignment-object}
\end{figure}

The Prompt KG is generated in real-time by instructing GPT-3.5 to extract the subject, relation, and object entities from the creature description found in the user's prompt. Our graph alignment process finds the Species KG that matches the relation and the relevant subject or object in the Prompt KG. First, we use semantic matching with vector embeddings to determine which characteristic name corresponds to the relation. For subject matching (see Figure~\ref{fig:KG-alignment-subject}), we return the list of characteristics corresponding to the relation for creatures where the scientific name or aliases match the subject. For object matching (see Figure~\ref{fig:KG-alignment-object}), we return a list of scientific names where the object matches one of the characteristics of the species corresponding to the relation. We use exact string matching for the subject and object to improve processing speed. 

If we cannot successfully align the KGs, we then attempt to perform exact string matching with normalized tokens from the description to the paragraphs saved from the Wikipedia pages for each species. This is fast but prone to false positives. If this also fails, we try to match the description from the prompt to vector embeddings of paragraphs from Wikipedia. This is much slower and frequently misses results.

Once the scientific names are identified, they are used in FathomGPT's SQL query generation component to retrieve data from the FathomNet database.

\subsection{SQL query generation}
\label{SubSec:Text2SQL}
The FathomNet database is a relational database with records of marine species found in global oceans. The data is stored in tables that contain images of marine life and also information about where individual creatures are located within the images. Each image is further described in other tables containing additional data, such as where and when the image was captured, and the temperature, oxygen levels, and water pressure at the time and location of the image capture. Thus, we use a text-to-SQL approach to search for information in the database. 

OpenAI provides APIs to fine-tune their existing model to better align its responses to our requirements~\cite{openai2024fine-tuning-api}. Fine-tuning improves model performance by training the model on data specific to our domain and formatted in the same way that we want the model to respond, enabling our model enables to more effectively integrate with the FathomNet database than it otherwise would be able to. For example, during the development process we found that the general knowledge available to the LLM would mask the more specific domain knowledge available in the FathomNet database. For example, FathomNet contains precise  longitudes and latitudes for marine regions, while asking the LLM about location information generally resolves to a single decimal. Other examples involve the non-fine-tuned model getting confused about the database schema and mismatching ids for data types that are unrelated, misinterpreting how the data should be ordered, and returning unexpected results that do not align with user prompts. (See our ablation study in Section~\ref{SubSec:Ablation1} for further details regarding the impact of fine-tuning.)

The SQL generation function is invoked by the prompt evaluator, which provides it with a prompt that has been augmented to include the previous context and the appropriate prompt type. We use different fine-tuned models for generating SQL queries based on the identified prompt type, which include similarity search prompts, visualization prompts, and prompts that output images, text, or tables. During development, we found that using three different models prevented the LLM from generating unnecessarily  complex SQL queries for even simple prompts.

Benchmarks on Text-to-SQL generations show that  prompt engineering methods can improve query generation~\cite{gao2023texttoSQL}. In-context learning is an effective approach that helps the model to understand the input prompt and its expected response. We provide the database structure to the LLM along with the foreign key and primary key information for the tables. We also include two demonstration examples in the prompt (based on the type of prompt the SQL query is created for). In cases where the SQL generation model produces errors, we provide the GPT-4 model with the problematic SQL and the error message and instruct it to produce new SQL that resolves the error~\cite{pourreza2024din}.


Once the SQL is generated, the FathomNet database will be queried and the results will be processed in a way that is appropriate to the prompt's expected output. The query itself is also available to the user for inspection.

\subsubsection{Textual outputs}
\label{SubSubSec:TextOutputs}
Some prompts will generate text outputs. For example, the user prompt ``Give me the name of the species found most frequently in the region Monterey Bay at a depth level lower than 5000m'' will successfully generate the following SQL:

\begin{verbatim}

SELECT TOP 1
    b.concept AS species,
    COUNT(*) AS frequency
FROM
    dbo.bounding_boxes AS b
    JOIN dbo.images AS i ON b.image_id = i.id
    JOIN dbo.marine_regions AS mr ON i.latitude 
    BETWEEN mr.min_latitude AND mr.max_latitude
    AND i.longitude BETWEEN mr.min_longitude 
    AND mr.max_longitude
WHERE
    mr.name = 'Monterey Bay'
    AND i.depth_meters < 5000
GROUP BY
    b.concept
ORDER BY
    frequency DESC;
\end{verbatim}

\noindent Here, the FathomNet database will return a response containing the columns ``species'' and ``frequency'' with a single row containing ``Strongylocentrotus fragilis'' as the species and ``11399'' as the frequency. Our fine-tuned model will incorporate these results into the generated response to produce the sentence: ``The most frequently found species in the region Monterey Bay at depth level lower than 5000m is Strongylocentrotus fragilis.''

Tabular output can be generated for some prompts, such as ``Generate me a list of top 20 species from the database that are found in pressure between 300 and 600 dbar''. The generated query returns a table of data, which is transformed by our fine-tuned model into a JSON object that is then rendered by our web interface.

We use the FathomNet API to generate taxonomic information and display it to the user. In a pre-processing step, we collect the taxonomic trees for each species available in FathomNet and create a dictionary that maps each taxon to its ancestor and children nodes. A user can write a prompt that asks for taxonomic information about a species, such as ``Show me the taxonomic tree of dinner plate jellyfish''. The result is presented as a hierarchy of text formatted to highlight  taxonomic relationships.

\subsubsection{Image outputs}
Most frequently, users craft prompts that request images of species. FathomGPT generates SQL queries that return image URLs from FathomNet, and then displays these images in a table in our web interface. In most cases, FathomNet images are associated with a set of scientific measurements collected at the time and location the image was captured, and because each image represents a particular species, we can also also retrieve the species taxonomy. When a user clicks on an image, we display a ``data card'' that highlights the bounding box of the species within the image and that provides the additional information alongside the image (see the example in Figure~\ref{fig:teaser}, top).



\begin{figure}[t]
    \centering
    \includegraphics[scale=0.215]{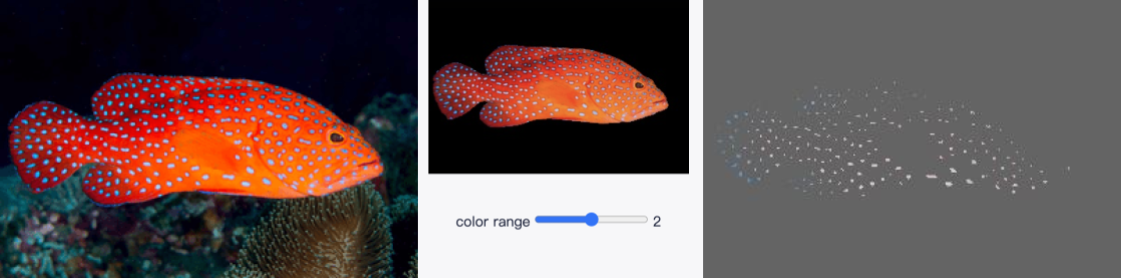}
    \includegraphics[scale=0.215]{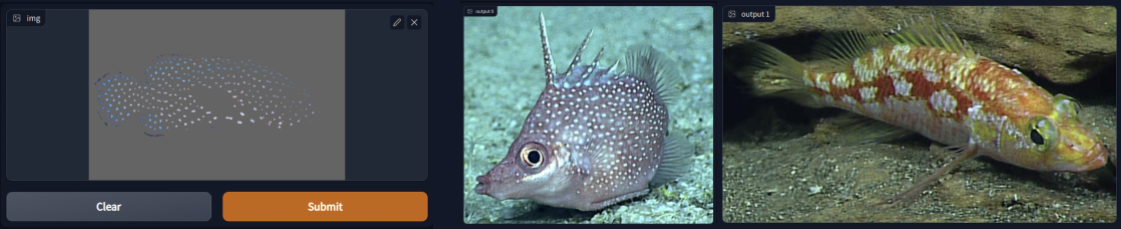}
    \caption{This figure outlines the pattern analysis interface. starting from the input image (Top left), pattern extraction of the main specimen (Top middle), extracted-pattern result (Top right), and finally searching through the database using the pattern (Bottom).}
    \label{fig:PatternAnalysis}
\end{figure}

Users can also upload their own image and perform a similarity search to find images in the database that match most closely. For example, a user could enter the prompt ``Find images similar to my input image at a depth greater than 2000 meters'' along with an input image. We extract feature vectors using a Vision Transformer model trained on ImageNet~\cite{Imagenet, dosovitskiy2020image} for each bounding box surrounding each species in each FathomNet image. (We found that using Vision Transformer for our similarity searches gave the best Top-1 Accuracy and Top-10 Recall Scores compared to the EfficientNet V2 Large, Densenet, and EfficientNet B7 models~\cite{tan2021efficientnetv2, tan2019efficientnet, huang2017densely}). Once the prompt evaluator determines that the user is requesting a similarity search, we extract the feature vectors from the input image via the Vision Transformer model, which then enables us to compute and rank the cosine similarity scores for the input image feature vector and the pre-computed feature vectors for the cropped images stored in the FathomNet database. Our SQL generation model can then utilize the cosine similarity scores to generate a SQL query that returns the relevant similar images.

\subsubsection{Visualization outputs}
Users can write prompts to generate custom interactive visualizations. FathomGPT uses the Plotly graphing library to produce these visualizations, which can include bar charts, line charts, heat maps, area charts, box plots, and scatter plots, among others (see Figure~\ref{fig:teaser}, bottom).

First, we instruct GPT-3.5-Turbo to generate an appropriate SQL query to retrieve the data necessary to populate the visualization and also to identify the data types that the visualization will ingest. We then simultaneously use the generated SQL to query FathomNet and at the same time instruct GPT-4 to generate Plotly code based on the identified data types and the user prompt. Once the SQL query returns data from FathomNet, it is passed into the generated Plotly code, which creates a Plotly object that is then displayed in our web interface. If the data types do not match, i.e., the query outputs do not align with the Plotly inputs, the visualization code will produce an error. We are able to catch the error and then regenerate the Plotly code using the query results themselves (see Section~\ref{subSubSec:Robustness}).




\begin{figure}[t]
    \centering
    \includegraphics[width=\columnwidth]{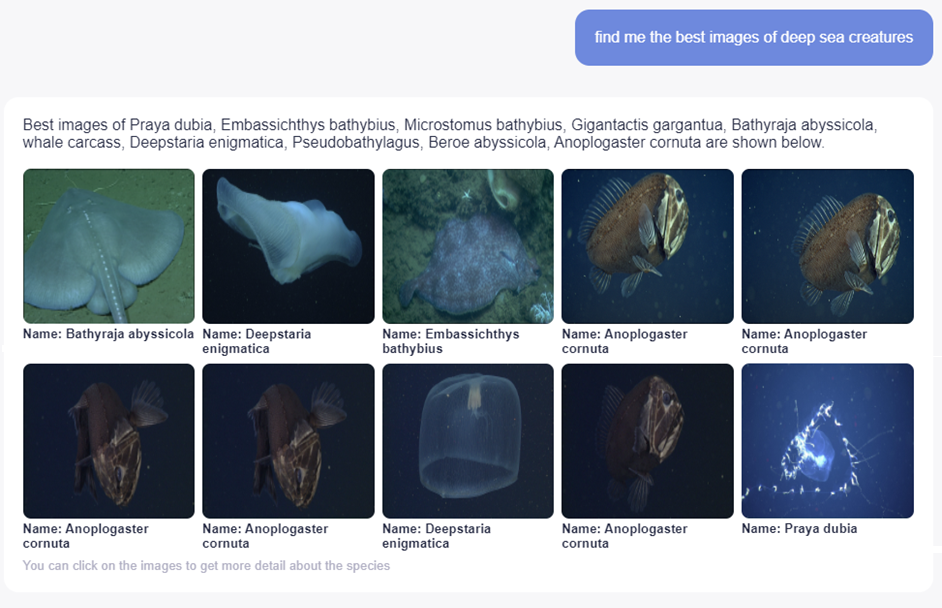} \\
    \caption{``Find the best images of deep sea creatures''}
    \label{fig:deepsea}
\end{figure}

During development, we noticed that GPT is able to produce effective complex code more readily than it can produce complex SQL. For example, if a user asks FathomGPT to generate a bar chart representing data that is split into multiple categories, this categorization can either be performed directly in the SQL or within the Plotly code. However, we found that the chances of the model generating an error were higher when generating SQL to categorize the data. At the same time, we want to ensure that the query does not return too much data, so we instruct the model to filter out data that is not required for the visualization, reducing both the amount of tokens and amount of data processing required by the Plotly code. That is, we leverage the abilities of the model to optimize the visualization generation process.

Users can also modify the Plotly code that is created simply by entering a follow-up prompt that asks FathomGPT to update the visualization. For example, a user might ask FathomGPT to ``Show a bar chart of population of species in Monterey Bay whose population is greater than 2000''. If the resulting visualization does not include a label for each bar, the user can simply ask FathomGPT to produce a new version of the visualization with a prompt such as ``Add the population number on top of each bar''.

\subsection{Pattern extraction and search}
\label{SubSec:PatternAnalysis}


Users can extract patterns from an uploaded image and then use them to search for FathomNet images containing similar patterns. We utilize a segmentation step before pattern extraction to ensure patterns that are taken from the uploaded image and are not influenced by other objects in the image. We use the Segment Anything model~\cite{Kirillov2023segment}, which enables us to rapidly generate accurate masks for objects without the need for additional fine-tuning. Segment Anything requires an initial point to begin the segmentation process, which is given in the form of a mouse click input to indicate which pattern needs to be extracted. The user is then presented with 3 masks of decreasing rigidity on the mask constraint (so that the different masks each cover a larger part of the pattern), as we found large variations in results depending on the input image. The masked image selected by the user is then utilized for further pattern extraction. The complete pattern analysis pipeline is showcased in Figure~\ref{fig:PatternAnalysis}.

\begin{figure}[t]
    \centering
    \includegraphics[width=\columnwidth]{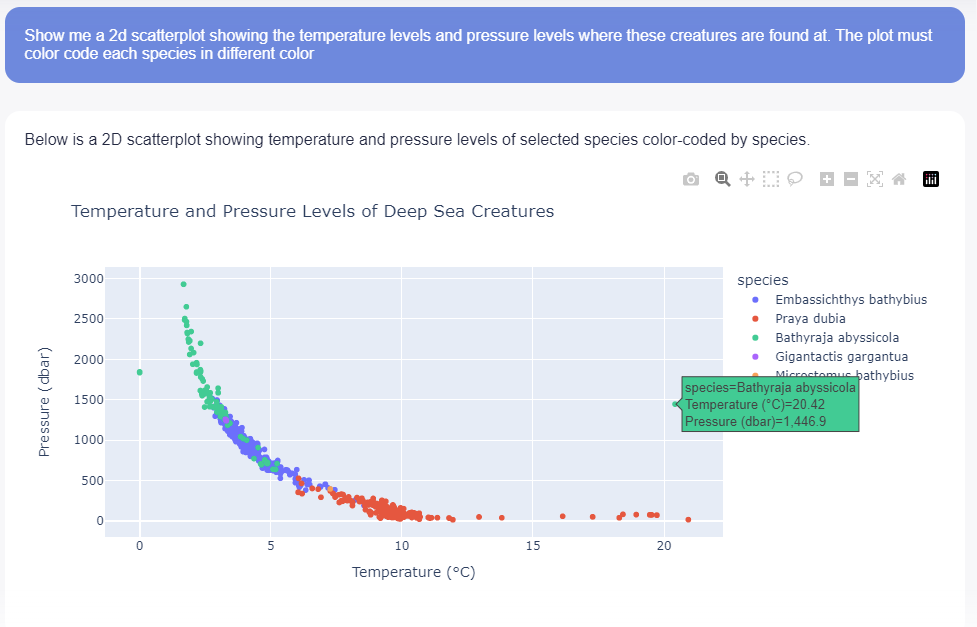} \\
    \caption{``Show me a 2d scatterplot showing the temperature levels and pressure levels where these creatures are found. The plot must color code each species in different color''}
    \label{fig:scatterplot}
\end{figure}

We utilize color differentiation to obtain complete patterns from the body of a specimen. Considering the physical phenomena of reflection, refraction, and scattering of light underwater, the same color might have different brightness levels. Hence, we use the HSV value to distinguish colors to obtain better results than using the RGB value. 
After the user selects a target HSV value, we mask the pixels with the same type of colors and generate a sub-image containing a pattern with the same color as the target color. Our pattern extraction interface enables a user to click on the target pixel in the image to extract the pattern corresponding to the color of this pixel. Moreover, users can adjust the color range value to determine the inclusivity of adjacent colors.


Once a pattern is extracted, it can then be used to search for similar patterns in the FathomNet database. In order to support better image similarity search on images using natural language, we search for not only the whole image but also parts of the image. We use EfficientNet V2~\cite{tan2021efficientnetv2} to generate feature vectors, striking a balance between speed and semantic richness. Search results are subsequently ranked according to the L2 norm and then presented to the user.

\section{Usage Scenarios} 
\label{Sec:UsageScenarios}


In this section, we present example usage scenarios describing how different users can engage with FathomGPT in order to retrieve and analyze data stored in the FathomNet database. All of the functionality demonstrated below (and more) is currently available via links available at our project website (\url{https://github.com/CreativeCodingLab/FathomGPT}), and we present additional examples in the Supplementary Video documentation accompanying this paper. 

\subsection{Scenario 1: Comparing the species Praya dubia and Bathyraja abyssicola}

FathomGPT can be used by ocean scientists to search for, visualize, and analyze data from the FathomNet database. In this scenario, the scientist wants to compare information about two different deep sea creatures, Praya dubia and Bathyraja abyssicola. They ask FathomGPT for their common names with the prompt ``What is the common name of Praya dubia?'' and learn that the two creatures are the giant siphonophore and the deepsea skate, respectively. Here we depict the free-form interrogation users can carry out during an exploratory session using FathomGPT.

The scientist first asks FathomGPT for some background information about the deep sea creatures available in FathomNet (Figure~\ref{fig:deepsea}). The scientist wants to understand the relationships between the temperature and pressure levels of these deep sea creatures, so she asks FathomGPT to generate a scatter plot comparing this information across both measurements. FathomGPT converts the natural language prompt to an SQL query, runs the query to obtain data from the FathomNet database, and then displays an interactive visualization (Figure~\ref{fig:scatterplot}). 

\begin{figure}[t]
    \centering
    \includegraphics[width=\columnwidth]{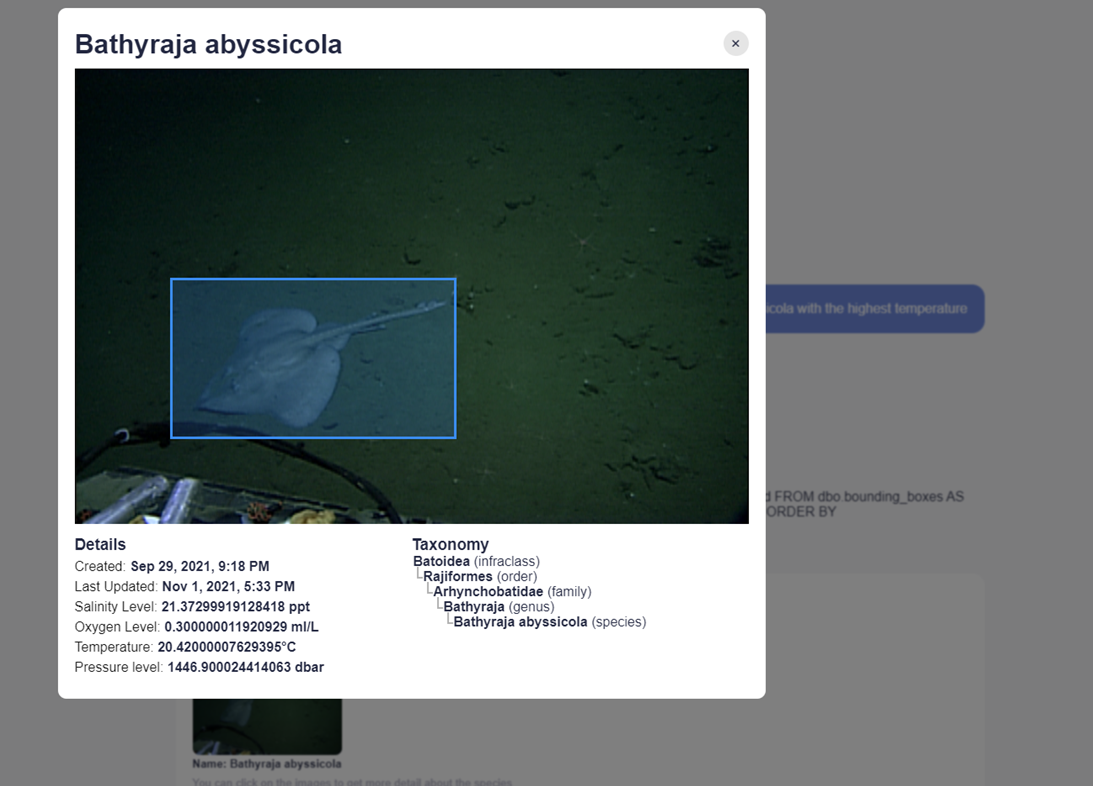} \\
    \caption{``Find the image of Bathyraja abyssicola with the highest temperature''}
    \label{fig:temperature}
\end{figure}

She immediately notices that the two species live in separate temperature and pressure levels. Specifically, she observes that one specimen of Bathyraja abyssicola is found at a temperature that was much higher than expected. She then investigates by asking FathomGPT about this outlier. FathomGPT then shows the scientist additional details, such as oxygen and salinity levels (Figure~\ref{fig:temperature}).

\begin{figure}[t]
    \centering
    \includegraphics[width=\columnwidth]{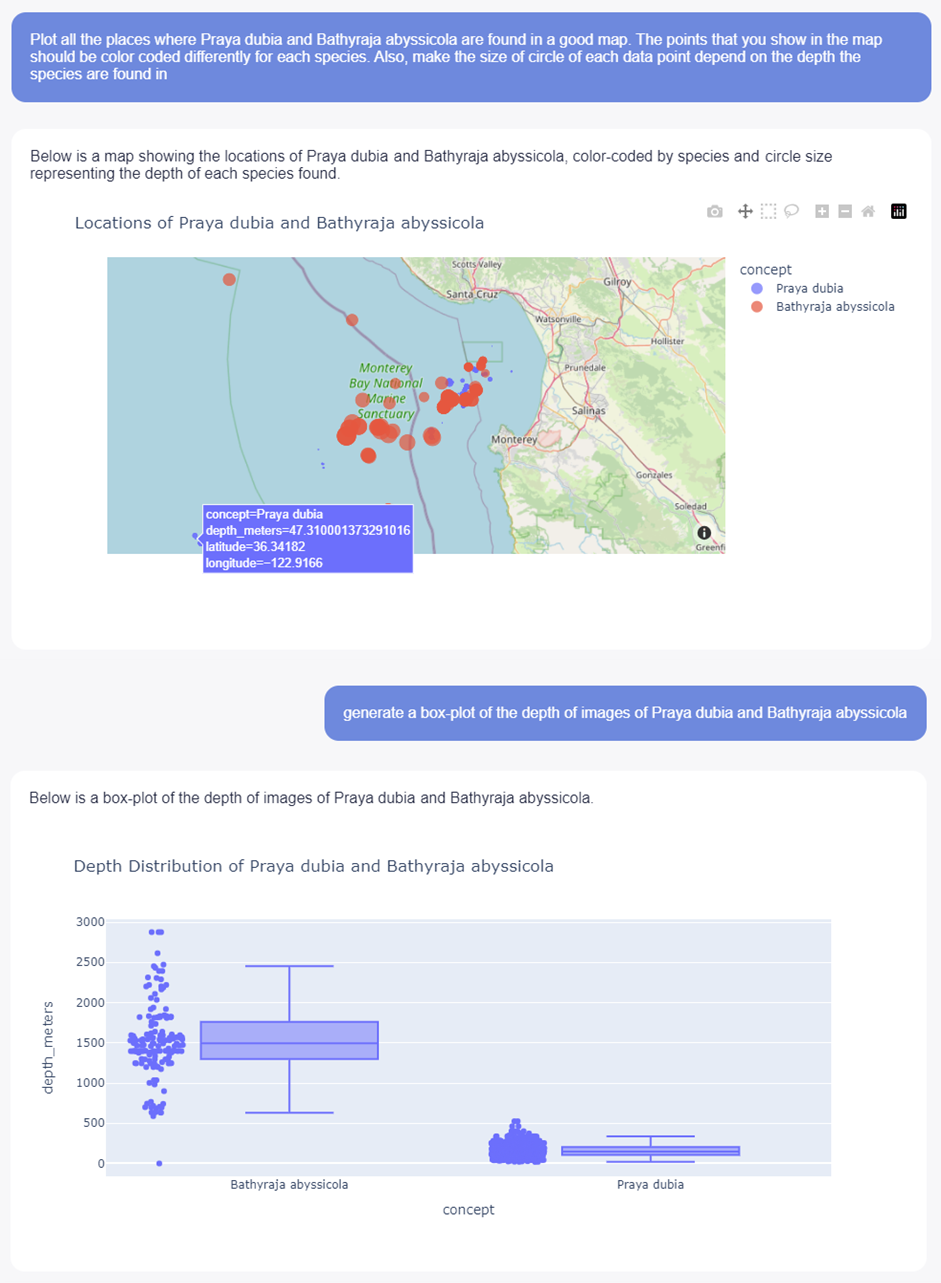} \\
    \caption{Top: ``Plot all the places where Praya dubia and Bathyraja abyssicola are found in a map. The points that you show on the map should be color-coded differently for each species. Also, make the size of the circle of each data point depending on the depth the species are found in''; Bottom: ``Generate a box-plot of the depth levels of images of Praya dubia and Bathyraja abyssicola''}
    \label{fig:depths}
\end{figure}

Since she is particularly interested in Praya dubia and Bathyraja abyssicola, she compares the regions and depths where the two creatures are found. She learns from FathomGPT that both species are found in the Monterey Bay region, but at different depths (Figure~\ref{fig:depths}). At this point, the scientist has learned something about both of these species and can continue using FathomGPT to ask additional questions as desired.

\subsection{Scenario 2: Locating regions with high populations of Aurelia aurita}
FathomGPT can also be used by ocean scientists to locate regions with a high concentration of a species and to inspect outliers. They can use the location information collected in previous ocean survey missions to make a determination about where to go to collect additional image samples.

\begin{figure}
    \centering
    \includegraphics[width=\columnwidth]{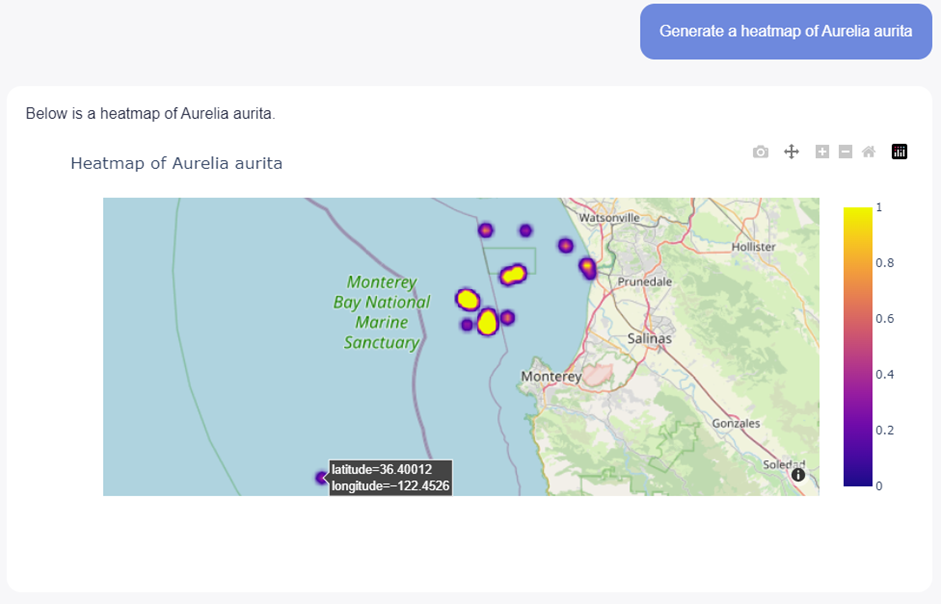} \\
    \caption{``Generate a heatmap of Aurelia aurita in Monterey Bay''}
    \label{fig:latlong}
\end{figure}

\begin{figure}
    \centering
    \includegraphics[width=\columnwidth]{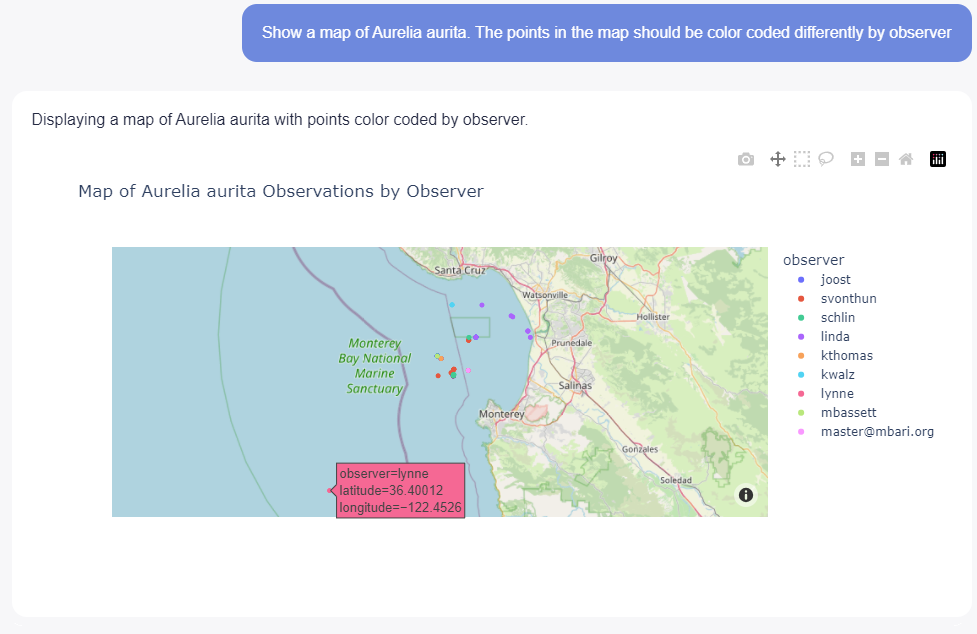} \\
    \caption{``Show a map of Aurelia aurita. The points in the map should be color-coded differently by observer''}
    \label{fig:observer}
\end{figure}

In this scenario, the ocean scientist wants to find the exact locations where Aurelia aurita (moon jellyfish) were previously observed. He asks FathomGPT to generate a heatmap of these observations (Figure~\ref{fig:latlong}). The scientist learns that many moon jellies have been observed in the area centered around (37°N, 122°W). He also noticed one sample that was found far away from the others. He plots the samples on a map color-coded by the image observer and notices that the outlier was collected during a different mission than the other samples, explaining why it was not located around the same area and providing a starting point for further investigation (Figure~\ref{fig:observer}). FathomGPT creates interactive charts, so the scientist can simply hover over a point to quickly find information about the observation. 

\subsection{Scenario 3: Identifying an unfamiliar species via a photo}
In addition to supporting scientific research tasks, FathomGPT can be used by citizen scientists and ocean enthusiasts as a tool to gather information about marine life. In this usage scenario, we provide an example of a user interested in identifying and learning more about a jellyfish that they were stung by on a beach in Monterey Bay.

\begin{figure}[t]
    \centering
    \includegraphics[width=\columnwidth]{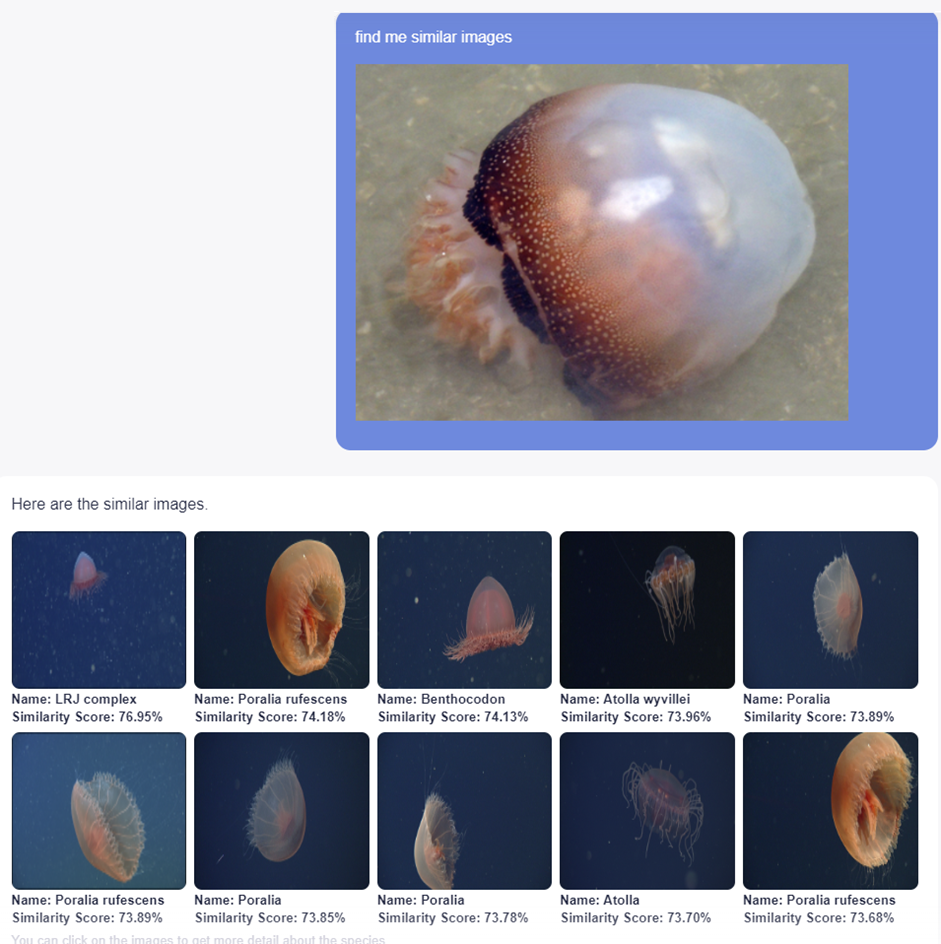} \\
    \caption{``Find similar images''}
    \label{fig:similar-jelly}
\end{figure}

First, the user uploads a photo\footnote{\url{https://inaturalist-open-data.s3.amazonaws.com/photos/1780063/medium.jpg}} of that jellyfish to FathomGPT to find the names of similar-looking creatures. The user sees that Benthocodon and Poralia rufescens look similar to the jellyfish that stung them (Figure~\ref{fig:similar-jelly}). The user further narrows down their question by determining which species they are most likely to find at the beach. The user prompts FathomGPT with questions such as "Are benthocodon found near the shore?" and learn that they are much more likely to encounter Poralia rufescens, since Benthocodon are usually found in deep sea environments instead.

\begin{figure}[t]
    \centering
    \includegraphics[width=\columnwidth]{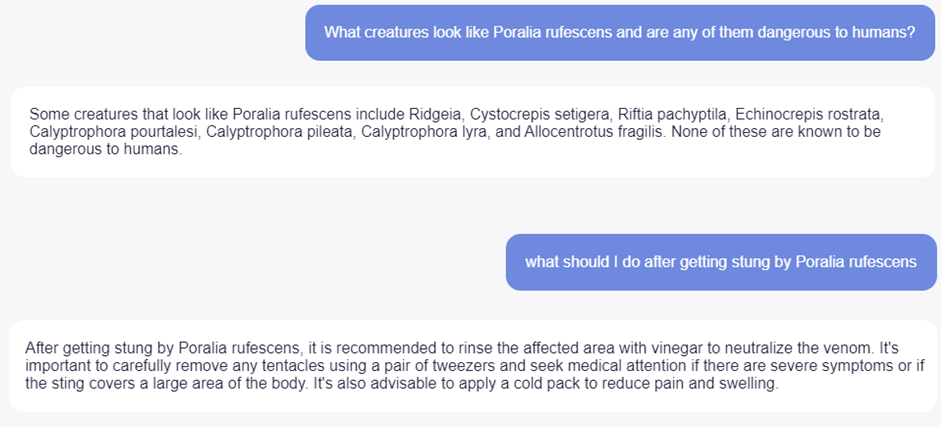} \\
    \caption{"Top: What creatures look like Poralia rufescens and are any of them dangerous to humans?"; Bottom: "What should I do after getting stung by Poralia rufescens?"}
    \label{fig:venomous}
\end{figure}

Out of curiosity, the user asks FathomGPT for additional species that look like Poralia rufescens to determine whether any of them are dangerous to humans and to ascertain whether or not there is a chance that they were stung by a similar-looking jellyfish. Now that the user is confident that the jellyfish is indeed Poralia rufescens, they ask FathomGPT about what they should do after they've been stung (Figure~\ref{fig:venomous}). FathomGPT helps the user identify which type of jellyfish stung them and further provides initial advice on how to handle the situation safely.

In each of the usage scenarios, we walk through an exploratory session that enables both researchers and citizen scientists to quickly ask complex questions using natural language. For more advanced users with a better understanding of the FathomNet database structure, the SQL queries generated by FathomGPT are also made available for inspection.

\section{System Evaluation} 
\label{Sec:Evaluation}

In this section, we present a series of studies evaluating the core components of the FathomGPT system. First, we analyze the latency during each substep of our system in order to isolate the factors contributing to latency reduction. Next, we present two ablation studies to show the impact that our fine-tuning approach and context passing strategy has on the responses to prompts.  Finally, we present an evaluation of our name resolution component in terms of its success at identifying scientific names in user prompts using FathomGPT's custom knowledge graph.

We recently introduced FathomGPT at a workshop for users of FathomNet that took place over two 4-hour sessions on March 14th, 2024. Attendees included 174 expert users with experience in ocean science and/or working with machine learning models in scientific contexts. The workshop attendees participated in multiple discussion sections that focused on potential applications, integration within existing research workflows, suitability for museum experiences, questions regarding programming details and about choices made when working with LLMs and machine learning models, and feature requests for future versions of FathomGPT. As part of this workshop, we invited participants to explore our system and to interact with FathomGPT using prompts related to their own research interests, logging 2977 individual prompts across 1816 interactive conversations.  

We used the anonymized data logged from the workshop to evaluate the performance of our system. Since there were many invalid queries from users testing how our system handles bad input as well as many duplicate queries from users inputting the prompts that are provided as examples directly within the interface, we filtered the data to obtain a unique set of reasonable queries. We additionally filtered out invalid queries that included playful attempts to ``hack'' the system, such as SQL injections or prompts meant to confuse the LLM, or that included nonsensical phrases. After filtering, 781 prompts remained, which we used as our main dataset for testing FathomGPT. (See the Supplementary Material for more details regarding our filtering methodology and deduplication strategy.) 

\subsection{Performance evaluation}

For our first study, we measure of the overall accuracy and query response time for this set of 781 prompts, and further analyze the latency during crucial subcomponents of FathomGPT in order to isolate the factors contributing to latency reduction. We also provide details regarding what component of the system caused any errors, providing insight each component's robustness. We provide reports for each FathomGPT component in the Supplemental Materials accompanying this paper, showing both the time they take and number of tokens they use. Additionally, we provide a set of examples in the Supplemental Materials that illustrate how a range of prompts from low to high prompt complexity are processed and passed from one function to another.



\subsubsection{Latency}
We measured the total time taken by FathomGPT to process each prompt, and found that the median time to process a prompt was 6.92 seconds (and that the minimum time taken was 1.45 seconds). In the Supplementary Materials, we analyze the latency at crucial points within the FathomGPT pipeline (see ~\autoref{fig:Pipeline}), which were designed specifically to greatly reduce the latency present in earlier iterations.

\subsubsection{Accuracy}
To evaluate the performance of the model, we labeled each response as being successful or as containing one or more errors. Successful responses occurred for 72.08\% of the prompts (563 out of 781). Each output was further categorized into 6 error categories. If the final output answered the user's prompt, the response was still be placed into these categories when we noted errors in the processing of the prompt. In the Supplementary Materials, we discuss these error categories and how they relate to the different FathomGPT components.

\subsubsection{Robustness}
\label{subSubSec:Robustness}
The name resolution, SQL generation, and Plotly code generation components of the system are able to resolve many errors that occur. For instance, as discussed in Section~\ref{SubSec:NameResolution}, the name resolution component contains multiple methods so that when one fails our system is able to fallback to another approach. Additionally, while running a generated SQL query, feedback is provided from the SQL server indicating any syntax errors, run-time errors, data type errors, or constraint violations errors. Our system is able to incorporate this feedback to regenerate a new SQL query, resolving the error. When evaluating the 781 prompts, we found 19 instances in which FathomGPT successfully fixed the initial errors (and only 1 where it failed to do so). Similarly, when running Plotly visualization code, exceptions generated by the visualization were caught and resolved. We found that in all 13 instances where there were compile time errors, FathomGPT successfully fixed the errors to produce working Plotly code.

\subsection{Ablation study \#1: Fine-tuning}
\label{SubSec:Ablation1}
As described in Section~\ref{SubSec:Text2SQL}, FathomGPT uses a custom fine-tuned model to generate SQL queries that retrieve image, text, and tabular results. To evaluate the performance and impact of our model, we performed an ablation study that compares our model to a non-fine-tuned model (i.e., one that only uses few shot prompting).  We categorized the output from each prompt in terms of 5 different error categories in order to evaluate its quality: Does not effectively utilize the database; Generates wrong data types; Generates unnecessarily large output; Generates overly complicated query; Generates a response that is missing data.

Overall, FathomGPT's fine-tuned model generated a correct response 14.54\% more often than the ablated version (252 vs. 220). Focusing on errors related specifically to SQL generation, FathomGPT’s model produced 46 errors versus 66 in the ablated version (out of 335 total).  During the labeling process, we found additional errors that did not match any of the expected categories, and these were placed in a catch-all ``Other errors'' category, and included various infrequent errors, such as generating wildly inaccurate SQL or generating empty strings, and errors due to missing constraints.  

For 5 out of 6 error categories, FathomGPT's fine-tuned model produced the same amount or fewer errors than the ablated version, and moreover we observed that fine-tuning generated better SQL than the ablated model that used only few shot prompting. For example, both versions return results successfully for the simple prompt ``find the best images of Sebastolobus alascanus''. However, the ablated version generated an SQL statement that inexplicably ordered results by timestamp rather than bounding box size, resulting in the presentation of lower quality images. As another example, one prompt asked our system to ``find images of different species of the superorder Octopodiformes''. The ablated version used fuzzy matching in the SQL rather than an exact match on the concept name. As a final example, we noticed that the ablated version repeatedly failed to generate correct SQL to retrieve similar images, whereas our fine tuned model had no issues handling prompts involving searching for images similar to a an input image provided by the user. 

Overall, fine-tuning the SQL generation model improves overall robustness and generates SQL queries that are more precise and that return more appropriate results. Please see the Supplementary Materials for more details regarding this study. 

\subsection{Ablation study \#2: Context}
\label{SubSec:Ablation2}
As discussed in Section~\ref{subsec:promptevaluator}, FathomGPT instructs the LLM to modify the current prompt to include relevant context from previous prompts during an conversation. We performed a second ablation study to evaluate this prompting strategy, comparing our prompt modification approach with an ablated version of our system in which the previous conversation is passed directly to the functions directly along with the current (unmodified) prompt. The data collected from the workshop contains 1816 conversations in total. After filtering out single prompt conversations, conversations that are not affected by context, and conversations that contain more than 50\% of the same prompts, we are left with 126 unique multi-prompt conversation containing 483 prompts, each of which was verified manually.

We found that FathomGPT’s prompting strategy generated a correct response for 7.27\% more queries versus the the ablated version (129 errors vs. 153 errors, out of 483 total), while average token used decreased by 15.67\% (10088 tokens vs. 11963 tokens). FathomGPT produced 46 results with an error due to the LLM not understanding the context, versus 56 for the ablated version. Some typical example prompts where FathomGPT performed better than the ablated version include the following:
\begin{itemize}

\item \textit{“Has it had any taxonomic revisions recently”}
The ablated version returned an answer directly via OpenAI’s API without using the context from the previous prompt mentioning the genus “anthothellidae”, while FathomGPT used the context correctly to produce the correct result.
\item \textit{“Can you show me some pictures?”} 
The ablated version retrieved images of the wrong species without using the context from previous prompt mentioning “anthothellidae”, while FathomGPT successfully made use of the context to retrieve the correct images.
\item \textit{“At what depth level is that species found at?”}
The ablated version did not use the species from the result of the previous prompt (“What species data is found in the deepest depth levels in the database?”), while FathomGPT used it correctly.
\item \textit{“That map does not look nice, draw a nicer looking map”} The user can prompt FathomGPT to update the visualization, but the ablated version returns the same visualization as the first prompt, whereas FathomGPT is able to successfully generate an improved visual representation.
\end{itemize}

The ablated model is less likely to correctly make sense of the context from previous prompts and results, while at the same time it requires more input tokens to process the prompt. This suggests that our prompting technique could improve the performance of LLM-based natural language interfaces. To the best of our knowledge, FathomGPT is the first system to include a prompting technique that instructs the LLM to generate modified prompts that incorporate specific contextual elements from the previous conversation in order to improve interactions with a scientific database.

\subsection{Evaluating KG name resolution}
\label{SubSec:EvalNameResolution}
We evaluated the accuracy of FathomGPT's name resolution functionality using a list of 185 prompts generated from FathomNet species or descriptions mentioned in user prompts collected from the workshop. These prompts ask for creatures based on their habitat, morphological features, color, or predator/prey relations.

We ran our name resolution for each prompt to obtain a list of at most 10 scientific names that match the description in the prompt. We then manually checked the results to determine how many of the scientific names fit the description. We then used GPT-4o directly and a vector embedding approach to resolve names in order to compare their accuracy to FathomGPT's name resolution, which features the custom knowledge graph described in Section~\ref{SubSec:NameResolution}. 




We passed each of the 185 prompts into FathomGPT's KG name resolution system, which returned correct results in most cases. It frequently gave results where all names are correct (63 out of 78 non-empty results), and on average, 92\% of the names in each result are correct. There were also many empty results, most (102 out of 107) due to prompts asking for predator/prey relations, which are not available for many of the species. For example, given a prompt asking for ``predators of hexanchus griseus'', KG name resolution correctly returns no results (as none are available in FathomNet). 



We also evaluated name resolution using GPT-4o directly to compare the accuracy of KG name resolution to state-of-the-art LLMs. We passed the same set of prompts along with a list of the 2000 FathomNet concepts to GPT-4o to limit the output to only species available in FathomNet. We found that GPT-4o hallucinates frequently, causing it to give inaccurate answers. There were only 17 out of 185 results where all names are correct, and on average only 56\% of the names in each result are correct (compared to 92\% for KG name resolution). There were no empty results, meaning it produced scientific names even when no species in FathomNet match the description.

When KG name resolution fails, FathomGPT will fallback to semantic matching with vector embeddings. Despite its simplicity, this method worked well for many prompts (on average, 72\% of the names in each result are correct), and has almost no latency since it requires no API calls. However, in addition to underperforming our KG name resolution, this approach is unable to handle prompts asking for creature-to-creature relations. For example, instead of understanding that the prompt is asking for predators of a species, it instead finds close matches of the species name instead. 

Comparing these three name resolution methods, FathomGPT's knowledge graph outperforms both vector embedding name resolution and the recently released GPT-4o. More details can be found in the Supplementary Materials.

\section{Conclusion \& Future Work}

In addition to providing us with a set of prompts with which to evaluate our system (see Section~\ref{Sec:Evaluation}), attendees of the FathomNet Workshop that took place on March 14th, 2024 engaged in multiple discussion sections with us that focused on potential applications, integration within existing research workflows, suitability for museum experiences, questions regarding programming details and about choices made when working with LLMs and machine learning models, and feature requests for future versions of FathomGPT. Feedback was overwhelmingly positive, and many participants told us that they eagerly look forward to the general release of FathomGPT and its integration into existing workflows that make use of the FathomNet image database. We especially appreciated hearing how researchers could accelerate some of their analysis tasks using our system, and the discussion with the machine learning experts gave us many ideas for continued work. 


For example, participants noted that in some cases our prompt evaluator model seemed sensitive to how a prompt is written. We plan to further fine-tune the model using different versions of a prompt to capture how a user might frame the same request. Participants were appreciative of the feedback that FathomGPT provides as it processes a prompt. They especially found that it was useful to see the generated SQL query, as it gave them assurance that the LLM was in fact retrieving data from the expected data tables. We will continue to explore additional ways to provide users insight into how FathomGPT processes results as it flows through the various intermediate steps. For instance, we are investigating integrating another LLM model into our system that observes the conversation, which could help us to evaluate if the user is provided with the expected result and could enable us to present more nuanced messages to the user.    

Participants also appreciated that our name resolution performed well, and were excited about its ability to resolve creature descriptions as well as names. Based on our discussions, we plan to expand our name resolution coverage to include more characteristics, such as species co-occurrence, in addition to the ones we already support. We will continue to collect more feedback from ocean scientists and other users to determine which characteristics could best serve their needs. Another issue that came up repeatedly is that users would like to be able to work with multiple descriptors in the same prompt (e.g., ``What are species that are orange and have tentacles?''). We will further explore advanced graph alignment techniques to handle more complex queries, such as the KG-Flex method~\cite{mckenna2023kgqa}. We also plan on scraping additional sources to obtain the species KGs, such as species descriptions from WoRMS and dichotomous keys~\cite{murguia2021taxonomic} found in ocean science textbooks.



Workshop participants were also intrigued with the pattern analysis capabilities included in FathomGPT. While our current pattern extraction approach works well for colorful species living in shallow water like tropical fish, the ocean contains many species with more subtle patterning or with transparent bodies. We continue to explore how we can extend our image processing tools and pattern extraction methods to support a much wider variety of marine life. Marine science has various intricate descriptors for different body structures based on shape, pose, color, and pattern that are used to identify species. A method to analyze images and present a dense natural language description of species features can help marine life enthusiasts learn more about different species and morphological features. While LLMs are adept at recognizing objects and elements in a picture, description tasks still elude these models to a large extent. Exploring ways to extend our pattern \textit{analysis} to facilitate pattern \textit{description} tasks could provide additional insight into ocean science images and beyond.


Finally, many users expressed interest in using FathomGPT as an interface for their own scientific datasets. Given that our system is highly dependent on having knowledge of the database schema and the data representations, this is not a straightforward task, but we are exploring ways both to use FathomGPT as an interface to facilitate ingesting new data into FathomNet and to make it easier to connect our system to other databases.

\begin{acks}
This work was supported by the National Science Foundation Convergence Accelerator Track E Phase I and II (ITE-2137977 and ITE-2230776). The authors gratefully acknowledge the FathomNet workshop participants who provided useful feedback. Special thanks to all the members of the Ocean Vision AI team, including Benjamin Woodward, Eric Orenstein, Geneviève Patterson, Kevin Barnard, Brian Schlining, Katy Croff Bell, and Susan Poulton.
\end{acks}

\balance
\bibliographystyle{ACM-Reference-Format}
\bibliography{FathomGPT}

\end{document}